\documentclass[aps,letterpaper,twocolumn,preprintnumbers,floatfix,superscriptaddress]{revtex4-1} 

\usepackage{amsmath}
\usepackage{amsfonts}
\usepackage{amssymb}
\usepackage{graphicx, rotating}
\usepackage{epstopdf}
\usepackage{epsfig}
\usepackage{latexsym}
\usepackage{color}
\usepackage[dvipsnames]{}
\usepackage{multirow}
\usepackage{xr}
\usepackage{booktabs}
\usepackage[table]{}
\usepackage{xcolor,colortbl}
\usepackage{float}

\usepackage{slashed}
\usepackage{hyperref}
\hypersetup{colorlinks, citecolor=bluscuro, linkcolor=bluscuro, urlcolor=bluscuro}
\definecolor{rossos}{cmyk}{0,1,1,0.55}
\definecolor{bluscuro}{rgb}{0.15, 0.2, .85}
\definecolor{bluchiaro}{cmyk}{1,.3,0.,0.1}
\definecolor{verdescuro}{rgb}{0.3,0.8,0.3}

\newcommand{\be}{\begin{equation}}
\newcommand{\ee}{\end{equation}}          
\newcommand{\bea}{\begin{eqnarray}}
\newcommand{\eea}{\end{eqnarray}}

\newcommand{\A}{{\cal A}}

\definecolor{newgreen}{RGB}{45,130,85}

\usepackage{bm}
\usepackage{mathrsfs}
\newcommand{\mat}[1]{\bm{\mathrm{#1}}}

\begin{document}

\widetext

\title{
Gravitational Compton Amplitude to All Orders in Perturbation Theory}

\author{Miguel Correia}
\affiliation{Department of Physics, McGill University, 3600 Rue University, Montr\'eal, H3A 2T8, QC Canada}
\author{Giulia Isabella}
\author{Anna M. Wolz}
\affiliation{Mani L. Bhaumik Institute for Theoretical Physics, Department of Physics and Astronomy, University of California Los Angeles, Los Angeles, CA 90095, USA}

\begin{abstract}
\noindent 
We show how amplitudes from the scattering of gravitational waves off compact objects in worldline effective field theory can be efficiently computed to arbitrary order in Newton’s constant $G$. Our approach solves an effective wave equation for the partial-wave amplitude in which Wilson coefficients (tidal Love numbers) enter through boundary conditions at short distances. We then perform the sum over partial waves to obtain the momentum-space Compton amplitude in terms of elliptic polylogarithms. We reproduce recent results through $\mathcal{O}(G^4)$ and obtain new predictions up to $\mathcal{O}(G^7)$, with tidal Love numbers first contributing at $\mathcal{O}(G^5)$. We find the first ultraviolet divergence in a classical gravitational amplitude at \(\mathcal O(G^7)\), showing that a pure point-particle description is not consistent in general relativity.  Matching to black hole perturbation theory, we show that static Love numbers vanish on-shell and predict subleading non-zero Schwarzschild black hole Love numbers.
\end{abstract}

\maketitle
\medskip

\noindent \textit{Introduction.} The observation of gravitational waves from black hole and neutron star binaries \cite{LIGOScientific:2016aoc,LIGOScientific:2017vwq} has motivated the development of rigorous theoretical descriptions of compact objects \cite{Buonanno:1998gg,Buonanno:2000ef,Goldberger:2004jt,Goldberger:2005cd,Goldberger:2009qd,Porto:2016pyg,Rothstein:2014sra,Kalin:2020mvi,Mogull:2020sak,Cheung:2018wkq,Kosower:2018adc,Buonanno:2022pgc,Goldberger:2022ebt}. Within worldline effective field theory (EFT), their finite-size response is encoded in Wilson coefficients, or tidal Love numbers \cite{Binnington:2009bb,Damour:2009vw,Goldberger:2004jt,Goldberger:2005cd}. These coefficients affect gravitational-wave observables \cite{Flanagan:2007ix,Hinderer:2007mb,Hinderer:2016eia} and provide a particularly sharp test of the nature of compact objects. For black holes, the static Love numbers are famously understood to vanish \cite{Binnington:2009bb,Damour:2009vw,Kol:2011vg,Porto:2016zng,Charalambous:2021kcz,Hui:2021vcv,Charalambous:2022rre,Ivanov:2022qqt,Goldberger:2022ebt}, while their frequency-dependent response contains nontrivial dissipative and running contributions \cite{Chia:2020yla,Perry:2023wmm,DeLuca:2024ufn,Chakraborty:2025wvs,Combaluzier--Szteinsznaider:2025eoc,Kosmopoulos:2025rfj,Chakraborty:2026dox,Solon:2026ubm,Kehagias:2026opd, Apostolidis:2026qsg}.

Scattering amplitudes provide an invariant framework for defining and extracting this response. In particular, the gravitational Compton amplitude describes the scattering of a gravitational wave off a massive object and directly connects the worldline EFT description to black hole perturbation theory (BHPT) \cite{Ivanov:2022qqt,Ivanov:2022hlo,Saketh:2023bul,Ivanov:2024sds,Saketh:2024juq,1973ApJ...185..635T,1973ApJ...185..649P,1974ApJ...193..443T,Mano:1996vt,Mano:1996gn,Mano:1996mf,Sasaki:2003xr,Bonelli:2022ten,Dodelson:2022yvn,Aminov:2023jve,Aminov:2024mul}. This connection has also played a central role in the modern amplitudes approach to compact-object dynamics, where Compton amplitudes have been constructed and matched to wave scattering in black hole backgrounds \cite{Bautista:2021wfy,Bautista:2022wjf,Ben-Shahar:2023djm, Bjerrum-Bohr:2025bqg,Bautista:2026fcp,Bjerrum-Bohr:2026fhs,Ivanov:2026icp,Bautista:2026qse,Brunello:2026rdk}. A basic tension between the two descriptions, however, is that BHPT naturally produces amplitudes partial wave by partial wave, whereas quantum field theory is naturally formulated in momentum space. Performing the infinite partial-wave sum while retaining the full
angular dependence is a classic problem in scattering theory, dating
back to the early days of wave mechanics
\cite{Watson:1918,Mott:1929,Taylor:1972pty,Newton:1982}.

A complementary route to gravitational-wave scattering was developed in ~\cite{Correia:2024jgr,Caron-Huot:2025tlq}. There the Feynman-diagram expansion was reorganized as the Born series of an effective wave equation. This made it possible to cleanly separate universal long-distance interactions from short-distance tidal effects and to compute partial-wave amplitudes to arbitrarily high orders in Newton's constant $G$. Applied to a scalar wave, the method yielded black hole response coefficients through $\mathcal{O}(G^7)$, far beyond the orders accessible through a direct diagrammatic expansion. Two important ingredients were nevertheless absent: the extension from a scalar probe to a genuine gravitational wave, including the recoil of the compact object, and the reconstruction of the momentum-space amplitude from the resulting partial waves.

In this Letter we supply both missing ingredients. We formulate the Born-series approach directly for graviton scattering from a dynamical compact object in worldline EFT, incorporating recoil and tidal interactions into an effective wave equation. Its solution determines the gravitational partial-wave amplitudes order by order in $G$, with the Wilson coefficients of the worldline theory entering as boundary conditions. We then show how the infinite partial-wave sum can itself be performed perturbatively, thereby reconstructing the momentum-space gravitational Compton amplitude. Remarkably, the resulting amplitudes can be expressed in terms of elliptic polylogarithms to all orders.

\vspace{8pt}

\noindent \textit{Effective Gravitational Wave Equation.}
The relevant action takes the form (we defer most technical details of this section to Appendix \ref{app:RWZ})
\begin{equation}
\label{eq:action}
    S=S_{\rm EH}+S_{\rm pp}+S_{\rm Love},
\end{equation}
where
\begin{equation}
\label{eq:EH}
    S_{\rm EH}
    =\frac{1}{16\pi G}\int d^d x\,\sqrt{-g}\,R, 
\end{equation}
is the Einstein-Hilbert action perturbed around a $d$-dimensional Schwarzschild background, $g_{\mu\nu}=\bar{g}_{\mu\nu}+2\mu^{\frac{4-d}{2}}M_{\rm Pl}^{-1}\,h_{\mu\nu}$, where $ M_{\rm Pl}^{-2}=8\pi G $ is the reduced Planck mass and
$\bar{g}_{\mu\nu}$ is given by
\begin{equation}
\label{eq:metric}
    ds^2 = - f dt^2 + f^{-1} dr^2 + r^2 d\Omega^2_{d-2}
\end{equation}
with
\begin{align}
\label{eq:fr}
  \!\! \!\!\! f(r) = 1 - {2 G M  \, n_d \,\mu^{4-d} \over r^{d-3}}, \quad\!\!\!\! n_d \equiv {4 \pi^{3 - d \over 2} \Gamma\big( {d -1 \over 2}\big) \over d - 2}.
\end{align} 
The metric $\bar{g}_{\mu\nu}$ is understood to be perturbatively expanded in Newton's constant $G$, whose units are kept as in four-dimensions via the introduction of the dimensionful scale $\mu$, where we will be making use of dimensional regularization $d = 4 - 2 \epsilon$ to regulate the UV divergences arising from the singularity $r = 0$.

Moreover, $S_\text{pp}$ and $S_\text{Love}$ are actions supported on the worldline of the compact object which is  moving with proper time $\tau$, velocity $u^\mu$ and coordinates $\bar{x}^\mu = x^\mu(\tau)$. The point-particle contribution $S_\text{pp}$ corresponds to the action for the geodesic $\bar{x}^\mu$ on the perturbed Schwarzschild background. It leads to a recoil term which, upon integrating out the worldline variation $\delta \bar{x}^\mu$, is given at quadratic order in $h_{\mu \nu}$  by \cite{Ivanov:2026icp}
\begin{equation}
\label{eq:pp}
    S_\text{pp} = (16\pi G M \mu^{4-d}) \int d\tau\, \delta \Gamma^\mu ( \bar x ) \frac{1}{\partial_\tau^2 }\delta \Gamma_\mu ( \bar x ) 
\end{equation}
with $\delta \Gamma^\mu(\bar x) = u^\alpha u^\beta \Gamma^\mu_{\alpha \beta}(\bar x)
=
u^\alpha u^\beta \bar{\nabla}_\alpha h^\mu_{\ \beta}(\bar x)
- \frac{1}{2}\,\bar{\nabla}^\mu \left( u^\alpha u^\beta h_{\alpha\beta}(\bar x) \right)$, where $\bar{\nabla}_\mu$ is the covariant derivative on the background metric $\bar{g}_{\mu \nu}$.

The finite-size response is encoded in the worldline EFT action through
multipoles of the Weyl tensor
\cite{Hui:2020xxx,Ivanov:2022hlo,Combaluzier--Szteinsznaider:2025eoc},
\begin{equation}
\begin{aligned}
S_{\rm Love}
&=
\sum_{\ell\geq2,n\geq0}
\frac{\mu^{4-d}}{\ell!}
\int d\tau
\Bigg[
\lambda_{n,\ell}^{(+)}\,
\mathcal E_{\text{avg},L}
(-r_sD_\tau)^n
\mathcal E_{\text{dif}}^{L}
\\
&\hspace{1.5cm}
+
\frac{\ell}{\ell+1}\,
\lambda_{n,\ell}^{(-)}\,
\mathcal B_{\text{avg},L|j}
(-r_sD_\tau)^n
\mathcal B_\text{dif}^{L|j}
\Bigg].
\end{aligned}
\label{eq:LoveAction}
\end{equation}
Here $D_\tau=u^\mu\nabla_\mu$ is the derivative along the worldline,
$r_s= 2 G M$ is the 4-dimensional Schwarzschild radius,
$L=i_1\cdots i_\ell$ denotes a symmetric trace-free (STF) spatial multi-index, and (avg, dif) indicate the Keldysh basis \cite{Keldysh:1964ud}. Via the projector orthogonal to the worldline, $P^\mu{}_\nu=\delta^\mu{}_\nu+u^\mu u_\nu$, with
    $\nabla^\perp_\mu=P_\mu{}^\nu\nabla_\nu$,  the electric tidal tensor and its multipoles are $\mathcal E_{\mu\nu}
    =
    C_{\mu\alpha\nu\beta}u^\alpha u^\beta,
$ and $
    \mathcal E_L
    =
    \nabla^\perp_{\langle i_1}\cdots
    \nabla^\perp_{i_{\ell-2}}
    \mathcal E_{i_{\ell-1}i_\ell\rangle}$, where $C_{\mu \alpha \nu \beta}$ is the Weyl tensor expanded to quadratic order in $h_{\mu \nu}$. For the magnetic sector we have
 $\mathcal B_{\mu\nu\rho}
    =
    P_\mu{}^\alpha P_\nu{}^\beta P_\rho{}^\gamma
    u^\sigma C_{\sigma\alpha\beta\gamma}$
which reduces to $\mathcal B_{ijk}=C_{0ijk}$ in the rest frame of the compact object. Its
multipoles are given by $
    \mathcal B_{L|j}
    =
    \nabla^\perp_{\langle i_1}\cdots
    \nabla^\perp_{i_{\ell-2}}
    \mathcal B_{i_{\ell-1}i_\ell\rangle j}$ and
   $\mathcal B_{(L|j)}=0$, where angle brackets denote STF projection. 
   We further collect the Wilson coefficients, or tidal Love numbers $\lambda^{(P)}_{n,\ell}$, into the response functions
\begin{equation}\label{eq:defLN}
    K_\ell^{(P)}(\omega) = {k^{(P)}_{\ell} \over \ell !} \sum_{n=0}^\infty
(i\omega r_s)^n\lambda_{n,\ell}^{(P)}
\end{equation}
where $P=-$ and $P=+$ denote the odd (magnetic) and even (electric) parity sectors,
respectively, and $k^{(+)}_{\ell} = 1$ and $k^{(-)}_{\ell} = \frac{\ell}{\ell+1}$ \footnote{For $d>4$ there is in addition a tensor-type tidal
sector, which decouples from the scalar and vector sectors and will
not be needed here.}.

Varying the action \eqref{eq:action} with respect to the metric
perturbation $h_\text{dif}^{\mu\nu}$ gives the equations of motion for the
gravitational wave in the worldline EFT, with Schwinger--Keldysh
doubling understood in the conservative actions \eqref{eq:EH} and
\eqref{eq:pp}. Decomposing $h_\text{avg}^{\mu\nu}$ into tensor spherical
harmonics and fixing the gauge as described in \cite{Kodama:2003jz,Hui:2020xxx}, the physical degrees of
freedom are described by radial master fields $\psi_\ell^P(r)$ (see Appendix \ref{app:RWZ}). The equations of motion read
\begin{equation}
\left[
\frac{d^2}{dr^2}
-\frac{(\ell-\epsilon)(\ell-\epsilon+1)}{r^2}
+\omega^2
\right]\psi_\ell^P(r)
=
V_\ell^P(r)\psi_\ell^P(r),
\label{eq:gravwaveeq}
\end{equation}
with the potential
\begin{equation}
    V_\ell^P
    =
    V_{\rm grav}^P
    +
    V_{\rm recoil}^P
    +
    V_{\rm Love}^P .
\label{eq:gravpotential}
\end{equation}
Here $V_{\rm grav}^P$ is the long-distance gravitational potential associated with the Schwarzschild background \eqref{eq:metric} and is
given in Eq. \eqref{eq:Vgrav-general}. Both $V_{\rm recoil}^P$ and $V_{\rm Love}^P$ are contact interactions supported on the worldline of the compact object. The recoil term vanishes in this gauge choice in dimensional regularization, as we show in Appendix~\ref{app:recoil}.  The Love number term, on the other hand, is proportional to the response function $V_\text{Love}^P \propto K^{(P)}_\ell(\omega) \,\delta(r)$ as given by Eq. \eqref{eq:VLoveRadial}. As in the scalar case
\cite{Caron-Huot:2025tlq}, $V_\text{Love}^P$ is most conveniently incorporated as
a boundary condition at the origin, as also noted in \cite{Combaluzier--Szteinsznaider:2025eoc}.

At short distances the two independent solutions behave as
\begin{equation}
    \psi_\ell^P(r \to 0)
    =
    \mu^{-\epsilon}B_{\rm reg}^P
    r^{1+\ell-\epsilon}
    +
    \frac{\mu^\epsilon B_{\rm irr}^P}
    {2\ell+1-2\epsilon}
    r^{\epsilon-\ell},
\label{eq:gravnearorigin}
\end{equation}
where the contribution of $V_\text{grav}^P$ drops out due to dimensional regularization.
 Integrating the wave equation through the contact
potential $V_\text{Love}^P$ therefore amounts to fixing the ratio
$B_{\rm irr}^P/B_{\rm reg}^P$ in terms of the corresponding
worldline response function, giving
\begin{equation}
\label{eq:BK}
  \!\!\!\!  {B_\text{irr}^P \over B_\text{reg}^P}\! =\! \frac{a_\ell^P}{M_{\rm Pl}^2}  \frac{2^{\ell-1}(\ell+d-2)}{\ell(\ell-1)\pi^{\frac{d-1}{2}}}\Gamma\big(\ell+\tfrac{d-1}{2}\big) K_\ell^{(P)}(\omega) 
\end{equation}
with $a_\ell^+ = \frac{(d-3)(\ell+d-3)\ell!}{d-2}$ and $a_\ell^- = \frac{(\ell+1)(\ell+1)!}{2\ell}$.

At large distances the solution takes the asymptotic form
\begin{equation}
    \psi^P_\ell(r\to\infty)
    =
    A^P_{\rm in}\,
    h^-_{\ell-\epsilon}(\omega r)
    +
    A^P_{\rm out}\,
    h^+_{\ell-\epsilon}(\omega r),
\label{eq:gravasymptotic}
\end{equation}
where $h^\pm$ are the Riccati-Hankel functions \cite{Taylor:1972pty},
from which we extract the partial-wave $S$-matrix
\begin{equation}
    S^P_\ell(\omega)
    =
    -\frac{A^P_{\rm out}}{A^P_{\rm in}}.
\label{eq:gravSmatrix}
\end{equation}
The effective wave equation \eqref{eq:gravwaveeq} can be solved perturbatively 
in $G$ for the partial-wave S-matrix via the Born series~\cite{Caron-Huot:2025tlq}. The resulting Born integrals can be reduced to harmonic polylogarithms (HPLs). It is more convenient to repackage the result in terms of the phase-shift $\delta_\ell^P$, defined via $S^P_\ell = e^{2 i \delta^P_\ell}$. In Appendix \ref{app:results} we list the phase-shift up to $\mathcal{O}(G^7)$ where ultraviolet and infrared divergences are regulated in dimensional regularization.

\vspace{8pt}

\noindent \textit{Scattering Amplitudes from Partial-Waves.} In this section, we reconstruct the momentum-space scattering amplitude from the partial-wave S-matrix. For each angular momentum \(\ell\) and parity sector \(P\) we define the partial-wave amplitude \(a_\ell^P(\omega)\) through
\begin{equation}\label{eq}S_\ell^P(\omega)=1+ia_\ell^P(\omega).\end{equation}
Following the polarization basis of~\cite{Ivanov:2026icp}, the four-dimensional amplitude can be decomposed as\begin{equation}\label{eq:ampHV}
    \A(z,\omega)= \A^H(z,\omega)H_{12}^2+\A^V(z,\omega)V_{12}^2,
\end{equation}
with $H_{12}$ and $V_{12}$ given in Appendix~\ref{app:partialwaves}. 
In general dimension there is an additional mixed structure proportional to \(H_{12}V_{12}\). This structure is a Gram determinant and vanishes identically in \(d=4\), leaving only the two terms displayed in Eq. \eqref{eq:ampHV}.

Each scalar coefficient \(\A^I\), with \(I={H,V}\), can then be expanded in partial waves: 
\begin{equation}\label{eq:PW}
    \A^I(z,\omega)= (2\omega)^{3-d}\sum_{\ell=2}^\infty n_\ell^{(d)}\sum_{P} a_\ell^P(\omega) \pi^{I}_{\ell,P}(z).
\end{equation}
Here \(n_\ell^{(d)}\) denotes the partial-wave normalization given in Eq.~\eqref{eq:nld} and \(\pi_{\ell,P}^I(z)\) the corresponding spin-2 tensor structures (see Eq.~\eqref{eq:H12V12} and the ancillary file of \cite{Ivanov:2026icp}). 

Inspecting the explicit expressions for \(a_\ell^P(\omega)\), we find two qualitatively different types of contributions:
\begin{equation}
    a_\ell^P(\omega)
    =
    a_{\ell,\mathrm{reg}}^P(\omega)
    +\sum_{\ell_\star}
    \delta_{\ell,\ell_\star}\,
    a_{\ell_\star}^P(\omega).
\end{equation}
All ultraviolet divergences are contained in terms supported at fixed
values $\ell=\ell_\star$. For these contributions the partial-wave sum
collapses and can be evaluated directly. Since the remaining
generic-$\ell$ contribution is UV finite, we may set $d=4$ before
performing the infinite sum \footnote{For generic scattering angle,
the partial-wave sum is also infrared finite. The individual partial
waves nevertheless contain the universal exponentiated infrared
divergence as an overall phase, which is inherited by the full
scattering amplitude.}.

Extracting the explicit Legendre polynomial from the partial-wave structures and collecting the remaining factors into \(\Pi_{\ell,P}^I(z,\omega)\),\(\bar\Pi_{\ell,P}^I(z,\omega)\), given in Eq. \eqref{eq:Pipm}, we reorganize the regular contribution as
\begin{align}\label{eq:PW-4D}
   &\A_{\mathrm{reg}}^I(z,\omega)
    =
    \sum_{\ell=2}^\infty\sum_{P}
    \left(\!\ell+\frac{1}{2}\!\right)
    P_\ell(z)\,\\\nonumber&\quad\quad\!\times\left[
   a_{\ell,\mathrm{reg}}^P(\omega)\,
    \Pi_{\ell,P}^I(z,\omega)
    + 
   a_{\ell-1,\mathrm{reg}}^P(\omega)\,
    \bar{\Pi}_{\ell,P}^I(z,\omega)\right]\!.
 \end{align}
The shifted $\ell$ in $a_{\ell-1,\mathrm{reg}}^P$ is necessary due to the derivatives of $P_\ell(z)$ appearing in the partial-wave structures $\pi_{\ell,P}^I(z)$ which can be written in terms of $P_{\ell+1}(z)$.
To perform the remaining infinite sum, we use the contour representation
\begin{equation}\label{eq:Legendre-contour}
    \left(\ell+\frac{1}{2}\right)P_\ell(z)
    =
    \frac{1}{2\pi i}
    \oint_{\Gamma}
    \frac{u^{2\ell+1}(1-u^4)}{2y^3}\,du,
\end{equation}
where
\begin{equation}\label{eq:y-definition}
    y\equiv\sqrt{1-2zu^2+u^4},
\end{equation}
and the original contour at infinity has been deformed onto the clockwise contour \(\Gamma\)
\begin{equation}
\includegraphics[width=0.6\linewidth]{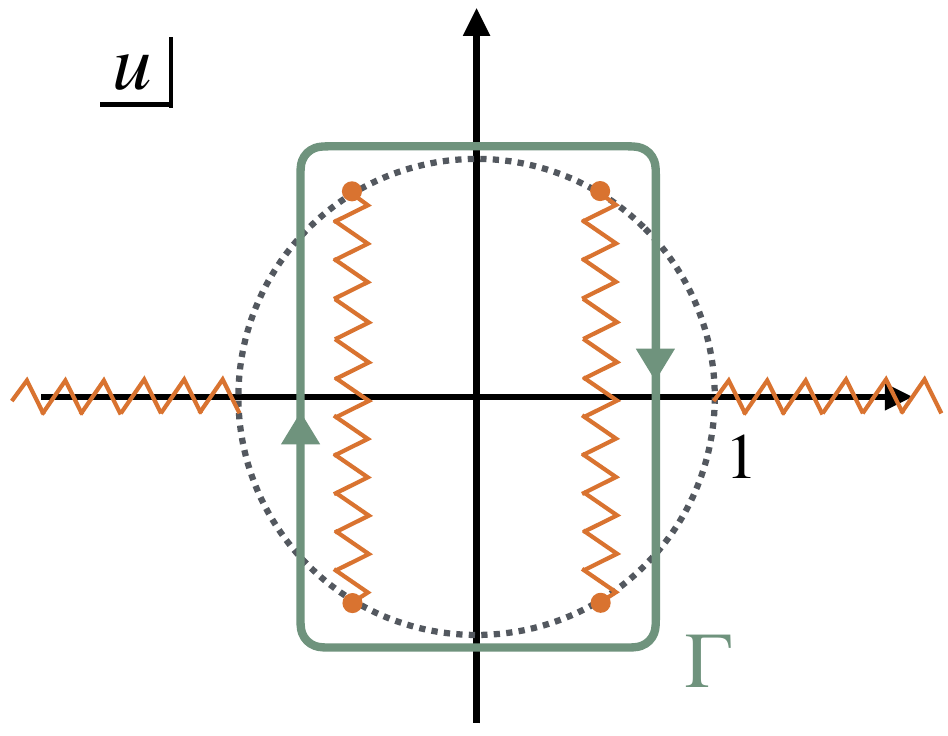}
    \label{fig:placeholder}
\end{equation}
 which surrounds the branch cuts of \(y\). Following this deformation, the partial-wave series converges on \(\Gamma\), so we may interchange the infinite sum and the contour integral and write
\begin{equation}
    \A_{\mathrm{reg}}^I(z,\omega)
    =
    \frac{1}{2\pi i}
    \oint_{\Gamma}
    \frac{u(1-u^4)}{2y^3}\,
    \mathcal G^I(u;z,\omega)\,du,
\end{equation}
where
\begin{equation}\label{eq:PWsumsG}
    \mathcal G^I(u;z,\omega)
    \equiv
    \sum_{\ell=2}^{\infty}\sum_P
    u^{2\ell}\,
    a_{\ell,\mathrm{reg}}^P(\omega)\,
    \Pi_{\ell,P}^I(z).
\end{equation}

After partial-fraction decomposition, the sums entering \(\mathcal G^I\) involve rational poles at integer or half-integer values of \(\ell\), polygamma functions, and products of these functions:
\begin{equation}
    \frac{1}{(\ell+m)^n},
    \qquad
    \frac{1}{\left(\ell+m+\frac12\right)^n},
    \qquad
    \psi^{(n)}(\ell).
\end{equation}
where $m$ is an integer.
These sums can be evaluated explicitly in terms of HPLs using the identities collected in Appendix~\ref{app:partialwaves}.

The resulting integrals are reduced to a basis of master integrals using integration-by-parts (IBP) identities. For each word \(\vec{g}\), there are six master integrals, which we collect into the vector \(\vec{I}_{\vec{g}}(x)\) where $x=\sqrt{{1-z \over 2}}$. An integral of a given weight can then be expressed as a linear combination of master integrals of the same or lower weight, with lower-weight contributions arising when derivatives act on the HPLs.

The six master integrals for a word $\vec{g}$ are defined as 
\begin{align}
\vec{I}_{\vec g}(x)
&\equiv \frac{1}{2\pi i}\oint
   \vec{\varpi}(u;x)\,H_{\vec g}(u),
   \label{eq:masterInts}
\\[5mm]
\vec{\varpi}(u;x)
&=\frac{du}{y}
\scalebox{0.88}{$
\begin{pmatrix}
 \tfrac{2}{\pi}\bigl[(1+u^2)K(x^2)-2u^2E(x^2)\bigr] \\[5pt]
 \tfrac{1}{\pi}\bigl[(u^2-1)K(1-x^2)-2u^2E(1-x^2)\bigr] \\[5pt]
 \dfrac{1}{x}\left(\dfrac{u^2+1}{u^2-1}\right) \\[5pt]
 u \\[5pt]
 \dfrac{1}{u} \\[5pt]
 x\left(\dfrac{1}{u-1}+\dfrac{1}{u+1}\right)
\end{pmatrix}$},
\notag
\end{align}
where $K$ and $E$ are the complete elliptic integrals of the first and second kind, respectively. The first three differential forms span the cohomology of the punctured torus, while the remaining three span that of the punctured sphere. $H_{\vec{g}}(u)$ is the HPL with word $\vec{g}$ using the conventions of \cite{Remiddi:1999ew}, with $H_\varnothing(u)=1$. The basis is chosen such that differentiation with respect to \(x\) lowers the weight by one
\begin{equation}\label{eq:masterIntsDE}
\begin{split}
    \frac{d}{dx}\vec{I}_{g_1,\vec{g}}(x)&= \mat{W}_{g_1}(x)\vec{I}_{\vec{g}}(x),\\
    \frac{d}{dx}\vec{I}_{\varnothing} (x) &= 0,
\end{split}
\end{equation}
where there are three possible matrices $\mat{W}_0$, $\mat{W}_1$, and $\mat{W}_{-1}$, given in Eq. \eqref{eq:Bmatrices}. The boundary conditions are fixed by evaluating Eq.~\eqref{eq:masterInts} in the backward limit at \(x=1\). At this point, the elliptic integration kernel simplifies, and the contour collapses onto the two poles at \(u=\pm i\). The resulting integrals therefore reduce to ordinary HPLs evaluated at \(i\). Using the identity \eqref{eq:ImReHPL}, we find, for any word \(\vec{g}\),
\begin{equation}\label{eq:generalBCs}
\vec{I}_{\vec{g}}(1)=\begin{pmatrix}\dfrac{4}{\pi}\operatorname{Im}H_{\vec{g}}(i)\\0\\0\\\operatorname{Re}H_{\vec{g}}(i)\\-\operatorname{Re}H_{\vec{g}}(i)\\-\operatorname{Re}H_{\vec{g}}(i)\end{pmatrix}
, \text{ with }
    \vec{I}_{\varnothing}(x)=\begin{pmatrix}
    0\\0\\0\\1\\-1\\-1
    \end{pmatrix}.
\end{equation}
Using the differential equation \eqref{eq:masterIntsDE}, the matrices \eqref{eq:Bmatrices}, and the boundary conditions \eqref{eq:generalBCs}, we can write the iterated solution
\begin{align}
\vec I_{\{g_n,...,g_1\}}(x)&= \vec{I}_{\{g_n,...,g_1\}}(1)\\\nonumber
&+\sum_{i=1}^{n}\bm{\mathcal W}_{\{g_n,\ldots,g_i\}}(x)\,\vec{I}_{\{g_{i-1},\ldots,g_1\}}(1),
\end{align}
where, for \(i=1\), the remaining word is understood to be empty. 
$\bm{\mathcal W}$ is defined recursively as iterated integrals over products of $\mat{W}$ matrices
\begin{align}
\bm{\mathcal W}_{\{g_n,\ldots,g_1\}}(x)
&=\!\!
\int_1^x dx_n\,
\mat W_{g_n}(x_n)\,
\bm{\mathcal W}_{\{g_{n-1},\ldots,g_1\}}(x_n),
\notag
\\[2mm]
\bm{\mathcal W}_{\varnothing}(x)
&=\mathbb I_6 ,
\label{eq:recursiveW}
\end{align}
where $g_i\in \{0,\pm1\}$. An explicit calculation of \(\vec{I}_{\{1\}}(x)\) is presented in App.~\ref{app:IntSolnEx}. The master integrals can also be evaluated numerically, either by direct integration along the finite contour or from their iterated-integral representation. We perform these numerical evaluations using \textsc{GiNaC} \cite{Bauer:2000cp,Vollinga:2004sn}.

\vspace{8pt}

\noindent \emph{Results.} 
After matching the partial-wave phase shift $\delta_\ell^P$ to the BHPT solution, we extract the magnetic and electric response through $\mathcal{O}(G^7)$ for $\ell=2$
\begin{align}\label{eq:BHLNs2}
   \frac{1}{M_{\rm Pl}^2}&K_2^{(-)}(\omega) =\frac{256i\pi}{135}   (GM)^6 \omega-\frac{512\pi}{135}  (GM)^7 \omega ^2 \nonumber \\[0.2cm]
    &\quad\times\left[\frac{1}{12 \epsilon }+\log (2GM\bar\mu)-\frac{377}{630}+\gamma_E \right],\\[0.2cm]
     \frac{1}{M_{\rm Pl}^2}&K_2^{(+)}(\omega) =\frac{128i\pi}{45}   (GM)^6 \omega-\frac{256\pi}{45}   (GM)^7 \omega ^2  \nonumber \\[0.2cm]
    &\quad\times\left[\frac{1}{12 \epsilon }+\log (2GM\bar\mu)-\frac{131}{168}+\gamma_E \right],
\end{align}
and $\ell=3$
\begin{equation}\label{eq:BHLNs3}
    \frac{1}{M_{\rm Pl}^2}K_3^{(-)}(\omega) =0, \qquad
     \frac{1}{M_{\rm Pl}^2}K_3^{(+)}(\omega) =0,
\end{equation}
with $\bar \mu^2=\mu^2 4\pi e^{-\gamma_E}$, verifying that the black hole static Love numbers are zero through $\mathcal{O}(G^7)$. 
Note that these Love numbers may be used as counterterms for ultraviolet divergences in the gravitational two-body problem, provided the corresponding scattering amplitudes are computed in the dim-reg scheme used in this work \footnote{
For instance, the shell-EFT scheme of
~\cite{Kosmopoulos:2025rfj,Solon:2026ubm} differs from the dimensional
regularization scheme used here.}. Upon renormalization we match the results of  \cite{Combaluzier--Szteinsznaider:2025eoc}.

The EFT exponentiated amplitude
\begin{equation}\label{eq:DeltaPW}
    \begin{split}
        \Delta^I(z,\omega)&=(2\omega)^{3-d}\sum_{\ell=2}^\infty n_\ell^{(d)}\sum_{P} 2\delta_{P}^{\ell}(\omega) \pi^I_{\ell,P}(z),
    \end{split}
\end{equation}
computed to $\mathcal{O}(G^7)$ and $\mathcal{O}(\epsilon^0)$ in $d=4-2\epsilon$, is given in Appendix~\ref{app:results} for $I=H$ and in the ancillary file for $I=V$. We additionally provide the corresponding full EFT amplitude $\A^I(z,\omega)$ (Eq. \eqref{eq:PW}) in the ancillary file.

We verify our results in several ways. First, we reproduce known results for $\Delta^I(z,\omega)$ through $\mathcal{O}(G^4)$ \cite{Bautista:2026qse,Bautista:2026fcp,Brunello:2026rdk,Ivanov:2026icp,Bjerrum-Bohr:2026fhs,Bjerrum-Bohr:2025bqg}. Second, we numerically compute the partial wave sum in Eq.~\eqref{eq:DeltaPW} from $\ell=2$ to $\ell=L_\text{max}$ and verify that the result converges with increasing $L_\text{max}$ to the amplitude $\Delta^I(z,\omega)$ with the master integrals evaluated numerically at several physical values of $(z,\omega)$. Third, we confirm that the EFT amplitudes $\Delta^I$ and $\A^I$ match the black hole amplitudes after substituting in the values for the black hole Love numbers given in Eqs. \eqref{eq:BHLNs2} and \eqref{eq:BHLNs3}.

\vspace{8pt}

\noindent \textit{Conclusion.} 
We have shown that the gravitational Compton amplitude can be expressed in terms of the master integrals $\vec{I}_{\vec{g}}$ belonging to the class of elliptic polylogarithms on the curve $y$ given in Eq. \eqref{eq:y-definition}, to all orders in perturbation theory. We have reproduced the literature through $\mathcal{O}(G^4)$ \cite{Bjerrum-Bohr:2025bqg,Bautista:2026fcp,Bjerrum-Bohr:2026fhs,Ivanov:2026icp,Bautista:2026qse,Brunello:2026rdk} and obtained new results through $\mathcal{O}(G^7)$. Our method is fully automatized and
going beyond $\mathcal{O}(G^7)$ is straightforward with additional computational power. We ran a conventional laptop over a few hours to produce the results in this work.

 Tidal Love numbers in the worldline EFT first contribute at $\mathcal{O}(G^5)$ and at $\mathcal{O}(G^7)$ we encounter the first ultraviolet divergence in a classical gravitational scattering amplitude. Its renormalization requires finite-size worldline operators, making manifest that a point-particle description is not self-consistent in general relativity. Matching to BHPT, we recover the vanishing of the static black hole Love numbers directly on shell and predict subleading dissipative and running Love numbers.

Several extensions are now within reach. One important direction is gravitational-wave scattering from rotating compact objects. Matching to solutions of the Teukolsky equation would determine the corresponding Kerr response coefficients and extend the tree-level results of ~\cite{Bautista:2021wfy,Bautista:2022wjf,Bautista:2023sdf,Saketh:2022wap,Ben-Shahar:2023djm,Guevara:2018wpp,Chiodaroli:2021eug,Cangemi:2023bpe,Bjerrum-Bohr:2023iey} to higher orders in $G$. We expect our framework to apply in a low-frequency expansion around Kerr, with the sum over spheroidal harmonics reducing perturbatively to a generalization of the partial-wave sums developed here.

Another natural application is gravitational-wave emission from multipolar sources such as compact binaries. The analogous problem has recently been explored for scalar radiation \cite{Chang:2026eti}, building on the Born-series framework of ~\cite{Correia:2024jgr,Caron-Huot:2025tlq}. Extending this construction to gravitons would provide a complementary route to gravitational waveforms and, possibly, to higher-point observables involving radiation in scattering processes \cite{Kosower:2018adc,Herderschee:2023fxh,
Brandhuber:2023hhy,Georgoudis:2025vkk}.

More broadly, our results provide an explicit perturbative map between partial-wave data and momentum-space scattering amplitudes for long-range interactions. The method should therefore have applications beyond classical gravity, including electromagnetic scattering in quantum mechanics and quantum field theory. Remarkably, the elliptic structures that arise after performing the partial-wave sum originate from comparatively simple meromorphic functions of angular momentum. This suggests that complex angular-momentum space may provide a useful organizing principle for classes of special functions and Feynman integrals that appear increasingly at high orders in perturbation theory \cite{Adams:2017ejb,Broedel:2018iwv,Becchetti:2025qlu}.

\noindent \emph{Note added. } The results of this work were first presented in \footnote{Seminar at Amplitudes 2026, Queen Mary University, London, June 29: ``Gravitational tidal matching and elliptic curves'', Giulia Isabella.}. While this work was being finalized, overlapping results at $\mathcal{O}(G^5)$ were published in \cite{Brunello:2026lzf}.

\vspace{2mm}

\noindent \emph{Acknowledgments.} We thank Zvi Bern, Simon Caron-Huot, Vasco Gonçalves, Misha Ivanov, Callum Jones, Julio Parra-Martinez, Julia Pasiecznik, Michael Ruf, Filipe Serrano, Mikhail Solon, and Zihan Zhou for useful discussions. We are particularly grateful to Simon Caron-Huot for valuable insights and collaboration at earlier stages of this work.
The work of  M.C. is supported by the National Science and Engineering Council of Canada (NSERC) and the Canada Research Chair program, reference number CRC-2022-00421. G.I. and A.M.W. are supported by the US Department of Energy under award number DE-SC0024224, the Sloan Foundation, the Mani L. Bhaumik Institute for Theoretical Physics, and the European Union (ERC grant GWSky/ 101167314). Views and opinions expressed are those of the authors and do not necessarily reflect those of the EU or the ERC Executive Agency. Neither the EU nor the granting authority can be held responsible for them. A.M.W. is supported by the NSF Graduate Research Fellowship under Grant No. DGE-2034835.

\bibliography{comptonG7} 

\clearpage
\mbox{}   

\appendix

\onecolumngrid
\section*{Supplemental Material}

\setlength{\parskip}{0.3em}

\section{More on the effective gravitational wave equation}
\label{app:RWZ}

\noindent Here we summarize the $d$-dimensional gravitational master equations
used in the main text, following ~\cite{Hui:2020xxx}. We consider
perturbations around the $d$-dimensional Schwarzschild background in \eqref{eq:metric}.
The metric perturbation can be decomposed into scalar,
transverse-vector, and transverse-traceless tensor harmonics on
$S^{d-2}$. Explicitly,
\begin{equation}
\begin{aligned}
h_{\mu\nu}
={}&
\sum_{\ell,m}
\begin{pmatrix}
 f(r) H_0(t,r) & H_1(t,r) & {\cal H}_0(t,r)\nabla_i \\
 * & f(r)^{-1}H_2(t,r) & {\cal H}_1(t,r)\nabla_i \\
 * & * &
 r^2\!\left[
 K(t,r)\gamma_{ij}
 +G(t,r)\nabla_{(i}\nabla_{j)_T}
 \right]
\end{pmatrix}
Y_\ell^m
\\[2mm]
&+
\sum_{\ell,m}
\begin{pmatrix}
0
&
0
&
h_0(t,r)Y_{ i}^{(T)m}{}_{\!\!\!\!\!\ell}
\\
*
&
0
&
h_1(t,r)Y_{ i}^{(T)m}{}_{\!\!\!\!\!\ell}
\\
*
&
*
&
r^2h_2(t,r)\nabla_{(i}Y_{j) }^{(T)m}{}_{\!\!\!\!\!\ell}
\end{pmatrix},
+
\sum_{\ell,m}
\begin{pmatrix}
0 & 0 & 0 \\
* & 0 & 0 \\
* & * & r^2h_T(t,r)Y_{ij}^{(TT)m}{}_{\!\!\!\!\!\ell}
\end{pmatrix}.
\end{aligned}
\label{eq:metric-decomposition}
\end{equation}
Here the entries denoted by $*$ are fixed by the entries across the diagonal because
$h_{\mu\nu}$ is symmetric. The harmonics $Y_\ell^m$ are scalar spherical
harmonics, $Y_i^{(T)m}$ are transverse vector harmonics, and
$Y_{ij}^{(TT)m}$ are transverse-traceless tensor harmonics.
The subscript $T$ denotes removal of the trace over the angular
indices. In the $d\to4$ limit, the scalar and transverse-vector sectors reduce
to the even (polar) and odd (axial) sectors, respectively, while the
independent tensor sector is absent. After fixing the gauge
\begin{equation}
    h_2={\cal H}_0=K=G=0
\label{eq:gaugeRWZ}
\end{equation}
and eliminating the constraint variables, the two physical sectors
are described by the Regge--Wheeler and Zerilli master fields
$\Psi_{\rm RW}$ and $\Psi_{\rm Z}$.

\subsection{Long-range gravitational field.}

\noindent The Regge-Wheeler and Zerilli equations with the pure long-range gravitational contribution are given by 
\begin{equation}
    \frac{d^2\Psi_{\rm RW}}{dr_*^2}
    +
    \left[
    \omega^2-V_{\rm RW}(r)
    \right]\Psi_{\rm RW}
    =0,
\label{eq:RWeq}
\end{equation}
and
\begin{equation}
    \frac{d^2\Psi_{\rm Z}}{dr_*^2}
    +
    \left[
    \omega^2-V_{\rm Z}(r)
    \right]\Psi_{\rm Z}
    =0,
\label{eq:Zeq}
\end{equation}
where the tortoise coordinate satisfies $  \frac{dr_*}{dr}=f^{-1}$.
The Regge--Wheeler potential is
\begin{equation}
V_{\rm RW}(r)
=
f\,\frac{(\ell+1)(\ell+d-4)}{r^2}
+
f^2\,\frac{(d-4)(d-6)}{4r^2}
-
ff'\,\frac{d+2}{2r}.
\label{eq:VRWappendix}
\end{equation}
The Zerilli potential is given by
\begin{align}
V_{\rm Z}(r)
={}&
\Big[
4(d-4)(d-2)^4 f^3
-8(d-2)^3(d-6)\ell(\ell+d-3)f^2
+4(d-2)^2(d-12)\ell^2(\ell+d-3)^2f
\nonumber\\
&\quad
+2(d-2)^3(d+2)r^3f'^3
-4(d-2)^2(d-6)\ell(\ell+d-3)r^2f'^2
-8(d-2)^2\ell^2(\ell+d-3)^2rf'
\nonumber\\
&\quad
+\big[d(d+10)-32\big](d-2)^3r^2ff'^2
+12(d-2)^5rf^2f'
-4(d-2)^3(3d-8)\ell(\ell+d-3)rff'
\nonumber\\
&\quad
+16(d-2)\ell^3(\ell+d-3)^3
\Big]
\frac{f(r)}
{4(d-2)r^2
\left[
2\ell(\ell+d-3)+(d-2)(rf'-2f)
\right]^2}.
\label{eq:VZappendix}
\end{align}
For $d=4$, Eqs.~\eqref{eq:RWeq} and \eqref{eq:Zeq} reduce to the
usual Regge--Wheeler and Zerilli equations \cite{Regge:1957td,Zerilli:1970se,Zerilli:1970wzz}.
Their higher-dimensional generalization was developed in
~\cite{Kodama:2003jz,Ishibashi:2003ap}. 

 To connect these equations with the representation \eqref{eq:gravwaveeq}
in the main text, we first rewrite them in terms of the Schwarzschild
radial coordinate $r$. Using $    \frac{d^2}{dr_*^2}   = f^2\frac{d^2}{dr^2}   +ff'\frac{d}{dr}$,
and defining
\begin{equation}
    \psi_\ell^P(r)
    =
    \sqrt{f(r)}\,\Psi_\ell^P(r),
    \qquad P=\pm,
\label{eq:master-redefinition}
\end{equation}
the first radial derivative is removed. Here $P=-$ denotes the
odd Regge--Wheeler sector and $P=+$ the even Zerilli sector.
The resulting equations take the form \eqref{eq:gravwaveeq} with the corresponding long-range gravitational potentials given by
\begin{equation}
V_{\ell,\rm grav}^P(r)
=
\omega^2\left(1-\frac{1}{f^2}\right)
+\frac{V_P(r)}{f^2}
-\frac{(\ell-\epsilon)(\ell-\epsilon+1)}{r^2}
+\frac{f''}{2f}
-\frac{f'^2}{4f^2},
\label{eq:Vgrav-general}
\end{equation}
where $V_-(r)=V_{\rm RW}(r)$ and   $V_+(r)=V_{\rm Z}(r)$. Both potentials vanish in the $G\to0$ limit and can be expanded
perturbatively in $G$ via the relation for $f(r)$ in \eqref{eq:fr},
yielding the long-range potentials that enter the Born series solution.

\subsection{Recoil potential}
\label{app:recoil}

\noindent Integrating out the worldline displacement gives the recoil action \cite{Ivanov:2026icp}, which in the frame where the compact object is at rest, $u^{\mu} = (1,\vec{0})$ and $\tau = t$, takes the form
\begin{align}
S_{\rm recoil}
=
16\pi GM\mu^{4-d}
\int d^d x\,
\delta^{d-1}(\bm{x})\,
\delta\Gamma^\mu
\frac{1}{\partial_t^2}
\delta\Gamma_\mu .
\label{eq:recoil-covariant}
\end{align}
Since this interaction is supported entirely at the position of the
compact object, it can affect the radial problem only through its
boundary condition at $r=0$. We derive the corresponding potential in the odd sector; the even-sector result follows analogously.
For the odd perturbations,
\begin{align}
\delta\Gamma_\mu^-
=
\left(
0,0,
\dot h_0-\frac{ff'}{2}h_1
\right)Y_i^{(T)m}{}_{\!\!\!\!\!\ell},
\end{align}
 hence the recoil action for a mode $\ell$ becomes
\begin{align}
S_{\rm recoil}^-
=
16\pi GM\mu^{4-d}
\int dt\,dr\,
\frac{\delta(r)}{r^2}
\left( \dot h_0-\frac{ff'}{2}h_1\right)\frac{1}{\partial_t^2}\left(\dot h_0-\frac{ff'}{2}h_1 \right).
\end{align}
Solving the bulk constraints and expressing the metric perturbations
directly in terms of the Born-series field \eqref{eq:master-redefinition} gives
\begin{align}
\label{eq:RWreconstruction}
h_0
=&
-q_\ell\,e^{-i\omega t}\,f^{\frac{1}{2}} r^{2-\frac{d}{2}}
\left(
r\partial_r
+\frac{d-2}{2}
-\frac{rf'}{2f}
\right)\psi_\ell^-,\qquad h_1
=
i\omega\, q_\ell\, e^{-i\omega t}\,f^{-\frac{3}{2}}r^{3-\frac{d}{2}}
\,
\psi_\ell^-
\\\nonumber
&
\text{with}\quad q_\ell
=
\frac{1}{\sqrt{2(\ell-1)(d+\ell-2)}} .
\end{align}
For $\omega\neq0$, the time derivatives cancel $1/\partial_t^2$, leaving a local contact operator:
\begin{align}
S_{\rm recoil}^-
=
16\pi GM\,q_\ell^2 \,\mu^{4-d}
\int \frac{d\omega}{2\pi}\,dr\,
f r^{2-d}\delta(r)
\left(
\mathcal D^-_\ell\psi_\ell^-
\right)^2,\quad \text{with}\quad \mathcal D^-_\ell
\equiv
r\partial_r+\frac{d-2}{2}-\frac{rf'}{f}.
\end{align}
The recoil potential is therefore conveniently specified by its
action on the wavefunction:
\begin{align}\label{eq:Vrecoil}
V_{\ell,\mathrm{recoil}}^-\psi_\ell^-=
- 32\pi GM\, q_\ell^2\,\mu^{4-d}
\Bigg\{
\left(
\frac{d-2}{2}-\frac{rf'}{f}
\right)
f r^{2-d}\delta(r)\,
\mathcal D_\ell^-\psi_\ell^-
-
\partial_r
\left[
f r^{3-d}\delta(r)\,
\mathcal D_\ell^-\psi_\ell^-
\right]
\Bigg\}.
\end{align}
It remains to determine whether this operator has any distributional
support at the origin. By replacing the wavefunction at $r\to0$ of Eq. \eqref{eq:gravnearorigin} and the metric \eqref{eq:fr} in \eqref{eq:Vrecoil}, we find that all terms appearing in the expression take the form $r^{a+b\epsilon}$ with $b\neq 0$. These vanish in dimensional regularization. It follows that $V_{\ell,\mathrm{recoil}}^-\psi_\ell^-=0$ in dimensional regularization.
The recoil interaction therefore does
not modify the boundary condition relating $B_{\rm irr}$ and
$B_{\rm reg}$. 
An equivalent derivation for the even sector shows that $V_{\ell,\mathrm{recoil}}^+\psi_\ell^+=0$ as well.

\subsection{Love number potential and boundary condition}
\label{app:love}

\noindent
We now derive the odd-parity Love-number potential appearing in
Eq.~\eqref{eq:gravpotential}. Fourier transforming the magnetic part
of the worldline action \eqref{eq:LoveAction} and using the response
function \eqref{eq:defLN} gives
\begin{equation}
S_{\rm Love}^-
=
\mu^{4-d}
\sum_{\ell\geq2}
\int\frac{d\omega}{2\pi}\,
K_\ell^{(-)}(\omega)\,
{\cal B}_{{\rm avg},L|j}(-\omega)
{\cal B}_{\rm dif}^{L|j}(\omega).
\label{eq:LoveOddFreq}
\end{equation}
We evaluate this interaction by projecting the metric onto a fixed
odd-parity $(\ell,m)$ mode. Since it is supported on the worldline,
gravitational corrections and repeated tidal insertions produce
noninteger powers of $r$ at the origin and vanish by analytic
continuation in dimensional regularization, as in the recoil
calculation. 

To convert the transverse vector harmonic in
Eq.~\eqref{eq:metric-decomposition} to Cartesian coordinates, we
introduce the constant mixed-symmetry tensor $c_{j|L}^{\ell m}$,
\begin{equation}
c_{j|L}^{\ell m}\,x^L
\frac{\partial x^j}{\partial\theta^A}
=
r^{\ell+1}Y_A^{(T)m}{}_{\!\!\ell},
\qquad
c_{(j|L)}^{\ell m}=0 ,
\label{eq:vectorSTF}
\end{equation}
where $c_{j|L}^{\ell m}$ is STF in $L=i_1\cdots i_\ell$ and
$x^L\equiv x^{i_1}\cdots x^{i_\ell}$. With the normalization of
Eq.~\eqref{eq:metric-decomposition} we have
\begin{equation}
c_{j|L}^{\ell m}c^{j|L,\ell m *}
=
\frac{2^{\ell-1}
\Gamma\!\left(\ell+\frac{d-1}{2}\right)}
{\pi^{\frac{d-1}{2}}\ell!}.
\label{eq:vectorSTFnorm}
\end{equation}
Taking the flat-space limit of the Regge--Wheeler reconstruction
\eqref{eq:RWreconstruction}, the Cartesian component of the regular
metric perturbation can be written directly in terms of the radial
field as
\begin{equation}
h_{0j}^{\ell m}
=
-\frac{q_\ell(d+\ell-2)}{\ell!}\,
c_{j|L}^{\ell m}x^L
\left.
\frac{d^\ell}{dr^\ell}
\left(
r^{\frac{2-d}{2}}\psi_\ell^-(r)
\right)
\right|_{r=0}
+\cdots .
\label{eq:h0Cartesian}
\end{equation}
Indeed, Eq.~\eqref{eq:gravnearorigin} implies that the derivative in
Eq.~\eqref{eq:h0Cartesian} equals
$\mu^{-\epsilon}\ell!B_{\rm reg}^-$ on the regular solution. At linear order in the canonically normalized perturbation, and using
the free linearized equations of motion, the magnetic multipole is
\begin{equation}
{\cal B}_{L|j}
=
\frac{\kappa}{2}
\left[
\partial_Lh_{0j}
-\partial_j\partial_{\langle L-1}h_{i_\ell\rangle0}
+\partial_0{\cal X}_{L|j}
\right],
\quad
{\cal X}_{L|j}
=
-\partial_{\langle L-1}h_{i_\ell\rangle j}
+\partial_j\partial_{\langle L-2}
h_{i_{\ell-1}i_\ell\rangle},
\label{eq:BlinearRW}
\end{equation}
where
$\partial_L$ denotes the STF projection of 
$\partial_{i_1}\cdots\partial_{i_\ell}$.
For the regular solution, Eq.~\eqref{eq:RWreconstruction} gives
$h_1\sim\omega r^{\ell+2}$, so that
${\cal X}_{L|j}\sim\omega r^2$ and its contribution vanishes at the
worldline. Moreover, the mixed symmetry in
Eq.~\eqref{eq:vectorSTF} implies
\begin{equation}
\left.
\partial_j\partial_{\langle L-1}h_{i_\ell\rangle0}
\right|_{\bm{x}=0}
=
-\frac{1}{\ell}
\left.
\partial_Lh_{0j}
\right|_{\bm{x}=0}.
\end{equation}
Using Eq.~\eqref{eq:h0Cartesian}, we therefore find
\begin{equation}
{\cal B}_{L|j}^{\ell m}
=
-\frac{\kappa}{2}\,
q_\ell\frac{\ell+1}{\ell}(d+\ell-2)\,
c_{j|L}^{\ell m}
\left.
\frac{d^\ell}{dr^\ell}
\left(
r^{\frac{2-d}{2}}\psi_\ell^-
\right)
\right|_{r=0}.
\label{eq:Bmaster}
\end{equation}
Substituting Eq.~\eqref{eq:Bmaster} into
Eq.~\eqref{eq:LoveOddFreq}, using
Eq.~\eqref{eq:vectorSTFnorm} and the value of $q_\ell$ in
Eq.~\eqref{eq:RWreconstruction}, gives the radial contact action
\begin{align}
S_{\rm Love}^-
={}&
\mu^{4-d}
\sum_{\ell,m}
g_\ell^-(\omega)
\frac{\mu^{2\epsilon}}{(\ell!)^2}
\int\frac{d\omega}{2\pi}\,dr\,\delta(r)
\,
\frac{d^\ell}{dr^\ell}
\left(
r^{\frac{2-d}{2}}
\psi_{{\rm avg},\ell m}^-(-\omega,r)
\right)
\frac{d^\ell}{dr^\ell}
\left(
r^{\frac{2-d}{2}}
\psi_{{\rm dif},\ell m}^-(\omega,r)
\right),
\label{eq:radialLoveAction}
\end{align}
where
\begin{equation}
g_\ell^-(\omega)
=
\frac{1}{M_{\rm Pl}^2}\,
\frac{2^{\ell-2}
\Gamma\!\left(\ell+\frac{d-1}{2}\right)}
{\pi^{\frac{d-1}{2}}}
\frac{(\ell+1)(\ell+1)!(d+\ell-2)}
{\ell^2(\ell-1)}
K_\ell^{(-)}(\omega).
\label{eq:gLoveB}
\end{equation}
Varying Eq.~\eqref{eq:radialLoveAction} with respect to the
difference field gives
\begin{align}
V_{{\rm Love},\ell}^-\psi_\ell^-
=
-\mu^{4-d}g_\ell^-(\omega)
\frac{\mu^{2\epsilon}(-1)^\ell}{(\ell!)^2}
r^{\frac{2-d}{2}}
\frac{d^\ell}{dr^\ell}
\left[
\delta(r)
\frac{d^\ell}{dr^\ell}
\left(
r^{\frac{2-d}{2}}\psi_\ell^-
\right)
\right].
\label{eq:VLoveRadial}
\end{align}
Thus the Love-number potential is a derivative contact interaction. We finally determine the corresponding boundary condition directly
from the distributional part of the wave equation. Using the
normalization of the irregular solution in
Eq.~\eqref{eq:gravnearorigin}, one has
\begin{equation}
\left[
\partial_r^2
-\frac{(\ell-\epsilon)(\ell-\epsilon+1)}{r^2}
\right]
\frac{\mu^\epsilon r^{\epsilon-\ell}}
{2\ell+1-2\epsilon}
=
-\frac{\mu^\epsilon(-1)^\ell}{\ell!}\,
r^{\frac{2-d}{2}}
\frac{d^\ell}{dr^\ell}\delta(r),
\label{eq:irrDistribution}
\end{equation}
On the other hand, acting with the contact potential
\eqref{eq:VLoveRadial} on the regular piece of
Eq.~\eqref{eq:gravnearorigin}, and matching both sides of the wave equation \eqref{eq:gravwaveeq} leads to the relation \eqref{eq:BK}. The even-parity derivation proceeds analogously; see also
~\cite{Apostolidis:2026qsg}.

\section{Partial Wave Summation}\label{app:partialwaves}
\noindent The partial wave normalization factor in Eq.~\eqref{eq:PW} is 
\begin{equation}\label{eq:nld}
    n_\ell^{(d)}= \frac{(4\pi)^{\frac{d-2}{2}}(d+2\ell-3)\Gamma(d+\ell-3)}{\Gamma\left(\frac{d-2}{2}\right)\Gamma(\ell+1)}.
\end{equation}
The amplitude can be decomposed in a basis of the functions (Eq.~\eqref{eq:ampHV}, see also \cite{Ivanov:2026icp})
\begin{equation}\label{eq:H12V12}
    H_{12}=\frac{1}{\omega^2}F_{\mu\nu}^{1}F^{2\mu\nu}, \qquad V_{12} = -\frac{4}{\omega^2}(F^1)_\mu^\nu (F^2)_{\nu\rho}u^\mu u^\rho +H_{12},
\end{equation}
which depend on the polarizations through 
\begin{equation}
    F^i_{\mu\nu} = i (p_{i\mu}\epsilon_{i\nu} -p_{i\nu}\epsilon_{i\mu}) .
\end{equation}
The spin-2 tensor structures can likewise be decomposed as
\begin{equation}\label{eq:pis}
\pi_{\ell,P}(z)
=
\pi_{\ell,P}^{H}(z)\,H_{12}^{\,2}
+
\pi_{\ell,P}^{V}(z)\,V_{12}^{\,2}
-\frac14\pi_{\ell,P}^{HV}(z)\,H_{12}V_{12},
\end{equation}
with $\pi_{\ell,P}^{I}(z)$ given in the ancillary file of \cite{Ivanov:2026icp}.

\vspace{.2cm}
The tensor structures $\pi_{\ell,P}^{I}$ contain Legendre polynomials $P_\ell(z)$ and its derivatives, which can be written in terms of $P_{\ell+1}(z)$. In order to make use of the identity in Eq.~\eqref{eq:Legendre-contour}, we separate out terms in Eq.~\eqref{eq:PW} (with $d=4$) that multiply $P_\ell(z)$ directly, denoted $\Pi_{\ell,P}^I$, and $P_{\ell+1}(z)$. Terms that multiply $P_{\ell+1}(z)$ are multiplied by $1=\frac{\ell+3/2}{\ell+3/2}$, and then $\ell$ is shifted to $\ell-1$, resulting in $\bar\Pi_{\ell,P}^I$, such that the overall coefficient $(\ell+1/2)P_\ell(z)$ is left over with the sum starting one higher. In $d=4$, there are two independent functions
\begin{align}
    \Pi_{\ell,+}^H(z,\omega) &= \frac{4\pi}{\omega}\frac{
  \ell\left(\ell^2+\ell-8\right)
  -2(\ell-1)(\ell+2)^2 z
  +(\ell+1)(\ell+2)^2 z^2
}{
  8(\ell-1)\ell(\ell+2)(1-z)^4
},\\
\bar\Pi_{\ell,+}^H(z,\omega) &= -\frac{4\pi}{\omega}\frac{
  \left(\ell-\frac{1}{2}\right)
  \left[2+\ell(\ell-1)(z-1)+z\right]
}{
  2(\ell-2)(\ell-1)
  \left(\ell+\frac{1}{2}\right)
  (\ell+1)(1-z)^4
},
\end{align}
with the others obtained from the relations
\begin{align}
     \Pi_{\ell,+}^H(z,\omega)&=-\Pi_{\ell,-}^H(z,\omega)=\Pi_{\ell,+}^V(-z,\omega)=\Pi_{\ell,-}^V(-z,\omega),\nonumber
     \\
     \bar\Pi_{\ell,+}^H(z,\omega)&=-\bar\Pi_{\ell,-}^H(z,\omega)=-\bar\Pi_{\ell,+}^V(-z,\omega)=-\bar\Pi_{\ell,-}^V(-z,\omega).
     \label{eq:Pipm}
\end{align}

\subsection{Identities for the partial-wave sum in terms of HPLs}
\noindent Here we collect the necessary identities to compute the partial wave sums $\mathcal G^I(u;z,\omega)$, defined in Eq.~\eqref{eq:PWsumsG}, in terms of HPLs. Note that, using the conventions of \cite{Remiddi:1999ew}, the letters of HPLs can be written by counting zeroes to the left of the letter $\pm1$; for example, $H_{3,-2,1}(u)=H_{0,0,1,0,-1,1}(u)$.

We write the polygamma function in terms of $\zeta$ functions and harmonic sums (for $\ell>0$)
\begin{equation}
    \psi^{(n)}(\ell+1)=(-1)^{n+1}n!\left(\zeta(n+1)-Z_{n+1}(\ell)\right), \qquad n\geq 1,
\end{equation}
and $\psi^{(0)}(\ell+1)=-\gamma_E+Z_{1}(\ell)$, where $Z_a(\ell)$ is the harmonic sum
\begin{align}
    Z_a(\ell)&=\sum_{n=1}^\ell \frac{1}{n^a},\quad a>0.
\end{align}
We use Stuffle relations to handle products of polygammas \cite{Remiddi:1999ew}, which generate iterated harmonic sums with strictly positive letters. For the rest of this section, let the vector $\vec{a}$ have only positive entries, and let $r=\text{Length }\vec{a}$. The iterated harmonic sums $Z_{\vec{a}}(\ell)$ obey
\begin{align}
    Z_\varnothing(\ell)&=1,\nonumber\\ 
    Z_{\vec{a}}(\ell)&=0,\quad 0\leq\ell< r,\\
\end{align}
with the recursive definition 
\begin{align}
    Z_{b,\vec{a}}(\ell)&=\sum_{n=r+1}^\ell \frac{1}{n^{b}}Z_{\vec{a}}(n-1),\quad b>0.
\end{align}
We use the following identities to compute sums over $\ell$ in terms of HPLs, with $n>0$,
\begin{equation}\label{eq:sumToHPL}
    \begin{split}
        &\sum_{\ell=0}^{\infty}\frac{u^{2\ell}}{\left(\ell+\frac12\right)^n}=\frac{2^{n-1}}{u}\left[H_n(u)-H_{n}(-u)\right],
        \\
        &\sum_{\ell=1}^{\infty}\frac{u^{2\ell}}{\ell^n}= H_n(u^2),
        \\
        &\sum_{\ell=1}^\infty u^{2\ell} Z_{\vec{a}}(\ell)= \frac{H_{\vec{a}}(u^2)}{1-u^2},
        \\
        &\sum_{\ell=0}^\infty \frac{u^{2\ell} Z_{\vec{a}}(\ell)}{\left(\ell+\frac12\right)^n} = \frac{2^{n-1+\sum_{j=1}^r a_j -r}}{u}\sum_{\epsilon_1,\ldots,\epsilon_r=\pm1}\left(\prod_{i=0}^{\lfloor (r-1)/2 \rfloor}\epsilon_{r-2i}\right)\left[H_{n,\vec{b}} (u)-H_{n,\vec{b}} (-u)\right],\\   &\qquad\qquad\qquad\qquad\qquad\qquad\qquad\qquad\qquad\,\text{with }\vec{b}=\left(\epsilon_1a_1,\epsilon_1\epsilon_2 a_2,\cdots,\left(\prod_{i=1}^{r}\epsilon_i\right)a_r\right),
        \\
        &\sum_{\ell=1}^{\infty} \frac{u^{2\ell} Z_{\vec{a}}(\ell-1)}{\ell^n} = H_{n,\vec{a}}(u^2).
    \end{split}
\end{equation}
The sums are convergent for $|u|<1$ and the identities can be verified by Taylor series expanding about $u=0$.
All sums appearing in $\mathcal G^I(u;z,\omega)$ can be written in terms of the above identities by using partial fraction decomposition in $\ell$, Stuffle identities, and by shifting the summation variable $\ell$ appropriately.

\subsection{Matrices for iterated master integral solution}
\noindent Using the shorthand notation
\begin{equation}
E=E(x^2),\qquad K=K(x^2),\qquad
\bar E=E(1-x^2),\qquad \bar K=K(1-x^2),
\end{equation}
and Legendre's relation 
\begin{equation}
    E\bar K+\bar E K-K\bar K=\frac{\pi}{2},
\end{equation}
the $\mat{W}$ matrices (Eq.~\eqref{eq:masterIntsDE}) are given by
\begin{equation}\label{eq:Bmatrices}
\begin{aligned}
\mat{W}_1(x)&=
\begin{pmatrix}
\mat{M}_1 & \mat{M}_2\\
\mat{M}_3 & \mat{M}_4
\end{pmatrix},
&
\mat{W}_{-1}(x)&=
\begin{pmatrix}
-\mat{M}_1 & \mat{M}_2\\
\mat{M}_3 & -\mat{M}_4
\end{pmatrix},
&
\mat{W}_0(x)&=
\begin{pmatrix}
\mat{M}_5 & 0_{3\times3}\\
0_{3\times3} & \mat{M}_6
\end{pmatrix},
\end{aligned}
\end{equation}
with
\begin{equation}
    \begin{split}
        \mat{M}_1&=\begin{pmatrix}
\dfrac{(1-x^2)\bar E K+2\bar K(K+Ex^2)}{\pi x(x^2-1)} & \dfrac{2K(E+2K)+2E(2E+3K)x^2}{\pi x(x^2-1)} & \dfrac{2(E+K)x^2}{\pi(x^2-1)}\\[10pt]
\dfrac{(1-x^2)(\bar E^2+3\bar E\bar K)+2\bar K^2}{2\pi x(1-x^2)} & \dfrac{\bar E(E+2K)(x^2-1)-\bar K(E+2K+Ex^2)}{\pi x(x^2-1)} & \dfrac{\bar K x^2}{\pi(1-x^2)}\\[10pt]
-\dfrac{\bar E}{4x^2} & \dfrac{E}{2x^2} & \dfrac{1}{2x}
\end{pmatrix},
\\[10pt]
\mat{M}_2&=\begin{pmatrix}
\dfrac{2(K+Ex^2)}{\pi x(x^2-1)} & 0 & \dfrac{2(E+K)}{\pi(x^2-1)}\\[10pt]
\dfrac{(1-x^2)\bar E+\bar K}{\pi x(1-x^2)} & 0 & \dfrac{\bar K}{\pi(1-x^2)}\\[10pt]
0 & 0 & \dfrac{1}{2x^3}
\end{pmatrix},
\\[10pt]
\mat{M}_3&=\begin{pmatrix}
-\dfrac{\bar E}{4x} & \dfrac{E}{2x} & \dfrac{x^2}{2(x^2-1)}\\[10pt]
-\dfrac{\bar E}{4x} & \dfrac{E}{2x} & \dfrac{x^2}{2(1-x^2)}\\[10pt]
0 & 0 & \dfrac{x}{2(1-x^2)}
\end{pmatrix},
\\[10pt]
\mat{M}_4&=\frac{1}{2x(1-x^2)}
\begin{pmatrix}
1-2x^2 & 0 & -x\\
1 & 0 & x\\
x & 0 & 1
\end{pmatrix}
,\\[10pt]
        \mat{M}_5&=\begin{pmatrix}
\dfrac{2\bar E K+2K\bar K-2\bar E K x^2+2E\bar K x^2}{\pi x(1-x^2)} & -\dfrac{4\left[K^2+E(E+2K)x^2\right]}{\pi x(x^2-1)} & 0\\[10pt]
\dfrac{(\bar E+\bar K)^2-\bar E(\bar E+2\bar K)x^2}{\pi x(x^2-1)} & \dfrac{2K(\bar E+\bar K)+2(-\bar E K+E\bar K)x^2}{\pi x(x^2-1)} & 0\\[10pt]
\dfrac{\bar E}{2x^2} & -\dfrac{E}{x^2} & 0
\end{pmatrix},\\[10pt]
\mat{M}_6&=\frac{1}{2x(1-x^2)}
\begin{pmatrix}
2x^2-1 & 1 & 0\\
-1 & 1-2x^2 & 0\\
-x & -x & 0
\end{pmatrix}.
    \end{split}
\end{equation}
The identities used to derive the boundary conditions of the master integrals in Eq.~\eqref{eq:generalBCs} are
\begin{equation}\label{eq:ImReHPL}
        \begin{split}
        \frac{1}{2\pi i}\oint \frac{H_{\vec{g}}(u)}{1+u^2}\,du&=\frac{H_{\vec{g}}(i)-H_{\vec{g}}(-i)}{2i}=\text{Im}H_{\vec{g}}(i),\\
\frac{1}{2\pi i}\oint \frac{u\,H_{\vec{g}}(u)}{1+u^2}\,du&=\frac{H_{\vec{g}}(i)+H_{\vec{g}}(-i)}{2}=\text{Re}H_{\vec{g}}(i).
    \end{split}
\end{equation}

\subsection{Example Master Integral Solution}\label{app:IntSolnEx}
\noindent We compute $\vec{I}_{\{1\}}(x)$ from the differential equation
\begin{equation}
\begin{split}
    \frac{d}{dx}\vec{I}_{\{1\}}(x)&=\mat{W}_1(x) \vec{I}_{\varnothing}(x)\\
    &=\begin{pmatrix}
    -\dfrac{2E(x^2)}{\pi(1+x)}-\dfrac{2\left(E(x^2)-K(x^2)\right)}{\pi x}\\[4pt]-\dfrac{E(1-x^2)}{\pi x}+\dfrac{K(1-x^2)-E(1-x^2)}{\pi(1+x)}\\[4pt]-\dfrac{1}{2x^3}\\[4pt]\dfrac{1}{2x}+\dfrac{1}{2(1+x)}\\[4pt]\dfrac{1}{2x}-\dfrac{1}{2(1+x)}\\[4pt]-\dfrac{1}{2x}+\dfrac{1}{2(1+x)}\end{pmatrix}.
\end{split}
\end{equation}
We integrate each term, and after matching at the boundary condition $x=1$  \eqref{eq:generalBCs} 
the solution is
\begin{equation}
\vec I_{\{1\}}(x)=\begin{pmatrix}
\dfrac{2+\pi-2E(x^2)-2(x-1)K(x^2)}{\pi}\\[5pt]\dfrac{-2E(1-x^2)+(1+x)K(1-x^2)}{\pi}\\[5pt]\dfrac14\left(\dfrac{1}{x^2}-1\right)\\[5pt]\dfrac12\left(\log\!\dfrac{x}{4}+\log(1+x)\right)\\[5pt]\dfrac12\left(\log 4 x-\log(1+x)\right)\\[5pt]\dfrac12\left(-\log x+\log(1+x)\right)
\end{pmatrix}.
\end{equation}

\section{Results}\label{app:results}
\subsection{Phase-shifts}
\noindent Using the following notation for the $\epsilon$-expanded Love numbers
\begin{align}
    \lambda^{(P)}_{\ell,n}(\epsilon)&=\frac{\lambda^{(P)}_{\ell,n}(\epsilon^{-1})}{\epsilon}+\lambda^{(P)}_{\ell,n}+\mathcal{O}(\epsilon^1),
\end{align} 
the EFT phase-shifts through $\mathcal{O}(G^7)$ are, for $\ell=2$,
\begin{align}\nonumber 
    \delta_2^+(\omega)&=2 GM \omega  \left[ -\frac{1}{2\epsilon_{\rm IR}}+\log \left(\frac{2 \omega }{\bar \mu_{\rm IR}}\right)-\frac{17}{12}+ \gamma_E \right]+\frac{107}{105} \pi  (GM \omega)^2\\[5pt]\nonumber
    &+(GM\omega)^3\left[\frac{428 }{105}\zeta_2-\frac{8 }{3}\zeta_3-\frac{259}{648}\right]+\frac{1695233 \pi }{1157625}(GM\omega)^4
    \label{eq:deltaPLUSl2G7}
    \\[5pt]
&+(GM\omega)^5\left[\frac{\lambda^{(+)}_{0,2} }{20 \pi  (GM)^5 M_{\rm Pl}^2}-\frac{3424}{525} \zeta_2^2+\frac{6780932 }{1157625}\zeta_2+\frac{91592 }{11025}\zeta_3+\frac{32 }{5}\zeta_5-\frac{403129}{1058400}\right]\\[5pt]\nonumber
&+(GM\omega)^6\left[ \frac{\lambda_{0,2}^{(+)}}{10 (GM)^5 M_{\rm Pl}^2}+\frac{i \lambda_{1,2}^{(+)}}{10 \pi  (GM)^5 M_{\rm Pl}^2}+\frac{756197385619 \pi }{140390971875}\right]\\[5pt]\nonumber
&+(GM\omega)^7\left[\frac{1}{\epsilon}\left(\frac{107 \lambda_{0,2}^{(+)}}{2100 \pi  (GM)^5 M_{\rm Pl}^2}-\frac{\lambda_{2,2}^{(+)}(\epsilon^{-1})}{5 \pi (GM)^5 M_{\rm Pl}^2}+\frac{32}{675}\right)\right.\\[5pt]\nonumber
&\quad\left.+\log \left(\frac{4 \omega ^2}{\bar\mu^2}\right) \left(-\frac{107 \lambda_{0,2}^{(+)}}{700 \pi  (GM)^5 M_{\rm Pl}^2}+\frac{\lambda_{2,2}^{(+)}(\epsilon^{-1})}{5 \pi  (GM)^5 M_{\rm Pl}^2}-\frac{224}{675}\right) \right.\\[5pt]\nonumber
&\left.\quad+\frac{675359 \lambda_{0,2}^{(+)}}{882000 \pi  (GM)^5 M_{\rm Pl}^2}+\frac{i \lambda_{1,2}^{(+)}}{5 (GM)^5 M_{\rm Pl}^2}-\frac{\lambda_{2,2}^{(+)}}{5 \pi  (GM)^5 M_{\rm Pl}^2}-\frac{9 \lambda_{2,2}^{(+)}(\epsilon^{-1})}{50 \pi  (GM)^5 M_{\rm Pl}^2}+\frac{2 \zeta_2 \lambda_{0,2}^{(+)}}{5 \pi  (GM)^5 M_{\rm Pl}^2}\right.\\[5pt]\nonumber
&\quad\left.+\frac{54784 }{3675}\zeta_2^3-\frac{143296 }{55125}\zeta_2^2+\frac{3024789542476 }{140390971875}\zeta_2+\frac{2902238896 }{121550625}\zeta_3-\frac{732736 }{11025}\zeta_5-\frac{128 }{7}\zeta_7+\frac{1017064145203}{420078960000}\right],
\end{align}

\begin{align}\nonumber
    \delta_2^{-}(\omega)& = 2GM\omega\left[-\frac{1}{2 \epsilon_{\rm IR}}+\log \left(\frac{2 \omega }{\bar \mu_{\rm IR}}\right)-\frac{5}{3}+\gamma_E \right]+ \frac{107 \pi }{105}(GM\omega)^2\\[5pt]\nonumber
    &+(GM\omega)^3\left[ \frac{428 }{105}\zeta_2-\frac{8 }{3}\zeta_3-\frac{29}{81}\right]+\frac{1695233 \pi }{1157625}(GM\omega)^4\\[5pt]\nonumber
&+(GM\omega)^5\left[\frac{\lambda_{0,2}^{(-)}}{20 \pi (GM)^5 M_{\rm Pl}^2}-\frac{ 3424}{525} \zeta_2^2+\frac{6780932}{1157625}\zeta_2+\frac{91592 }{11025}\zeta_3+\frac{32 }{5}\zeta_5-\frac{25609}{66150}\right]
\label{eq:deltaMINUSl2G7}
\\[5pt]
&+(GM\omega)^6\left[\frac{\lambda_{0,2}^{(-)} }{10 (GM)^5 M_{\rm Pl}^2}+\frac{i \lambda^{(-)}_{1,2}}{10 \pi  (GM)^5 M_{\rm Pl}^2}+\frac{756197385619 \pi }{140390971875} \right]\\[5pt]\nonumber
&+(GM\omega)^7\left[\frac{1}{\epsilon}\left(\frac{107 \lambda_{0,2}^{(-)}}{2100 \pi  (GM)^5 M_{\rm Pl}^2}-\frac{\lambda_{2,2}^{(-)}(\epsilon^{-1})}{5 \pi  (GM)^5 M_{\rm Pl}^2}+\frac{32}{675} \right)\right.\\[5pt]\nonumber
&\quad+\log \left(\frac{4 \omega ^2}{\bar\mu^2}\right) \left(-\frac{107 \lambda_{0,2}^{(-)}}{700 \pi  (GM)^5 M_{\rm Pl}^2}+\frac{\lambda_{2,2}^{(-)}(\epsilon^{-1})}{5 \pi (GM)^5 M_{\rm Pl}^2}-\frac{224}{675}\right)\\[5pt]\nonumber
&\quad +\frac{392717 \lambda_{0,2}^{(-)}}{441000 \pi  (GM)^5 M_{\rm Pl}^2}+\frac{i \lambda_{1,2}^{(-)} (1,2)}{5 (GM)^5 M_{\rm Pl}^2}-\frac{\lambda_{2,2}}{5 \pi  (GM)^5 M_{\rm Pl}^2}-\frac{77 \lambda_{2,2}^{(-)}(\epsilon^{-1})}{150 \pi  (GM)^5 M_{\rm Pl}^2}+\frac{2 \zeta_2 \lambda_{0,2}^{(-)}}{5 \pi  (GM)^5 M_{\rm Pl}^2}\\[5pt]\nonumber
&\quad \left.+\frac{54784}{3675}\zeta_2^3-\frac{143296 }{55125}\zeta_2^2+\frac{3024789542476 }{140390971875}\zeta_2+\frac{2902238896 }{121550625}\zeta_3-\frac{732736 }{11025}\zeta_5-\frac{128}{7}\zeta_7+\frac{34189466369}{13127467500}\right].
\end{align}
with $\bar \mu_\text{IR}^2=\mu_\text{IR}^24\pi e^{\gamma_E-1}$.
Then for $\ell\geq3$ they are given by the general formula (here, we set the $\ell=3$ static Love numbers to zero, but they are included in the ancillary file)
\begin{align}
    \delta_\ell^P(\omega) &= 2 GM \omega  \left(-\psi_P^{(0)}(\ell)+\frac{1}{2 \epsilon_{\rm IR}}+\log \left(\frac{2\omega }{\bar\mu_{\rm IR}}\right)-\frac{1}{2}\right)+ (GM\omega)^2\pi k_{1}^P\\\nonumber
    &+(GM\omega)^3\left[\frac{4}{3}\psi_P^{(2)} +4k_1^P \psi_P^{(1)}+k_2^P\right]+ (GM\omega)^4\pi k_3^P\\\nonumber
    &+(GM\omega)^5\left[  - \frac{4}{15}\psi_P^{(4)}(\ell)- \frac{8}{3}k_1^P\psi_P^{(3)}(\ell) - 4(k_1^P)^2\psi_P^{(2)}(\ell) +4k_3^P \psi_P^{(1)}(\ell)+ k_4^P\right]\\\nonumber
    &+(GM\omega)^6\pi k_5^P+(GM\omega)^7 \left[ \frac{8}{315}\psi_P^{(6)}(\ell)+\frac{8}{15}k_1^P\psi_P^{(5)}(\ell)+\frac{8}{3}(k_1^P)^2\psi_P^{(4)}(\ell)\right.\\\nonumber
   & \quad\left.+\frac{8}{3}\left[(k_1^P)^3-k_3^P\right]
\psi_P^{(3)}(\ell)-8k_1^P k_3^P \psi_P^{(2)}(\ell)+4k_5^P\psi_P^{(1)}(\ell) + k_6^P
\right],
\end{align}
with
\begin{align}
    k_{1}^P &= \frac{1}{32} \left[-256 \left(\frac{1}{\ell+1}+\frac{1}{\ell}\right)+225 \left(\frac{1}{2 \ell+3}+\frac{1}{2 \ell-1}\right)+\frac{694}{2 \ell+1}\right],\\[10pt]
    k_2^P &= 16P \left(\frac{1}{\ell^3}-\frac{1}{(\ell+1)^3}\right) +(84 P-4)\left(\frac{1}{\ell}-\frac{1}{\ell+1}\right) \\\nonumber&+\left(96 P-\frac{17}{2}\right)\left(\frac{1}{2 \ell+3}-\frac{1}{2 \ell-1}\right) +4 P\left(\frac{1}{\ell-1}-\frac{1}{\ell+2}\right) ,\\
    k_3^P & = 64 \left(\frac{1}{\ell^3}+\frac{1}{(\ell+1)^3}\right)-144 \left(\frac{1}{\ell^2}-\frac{1}{(\ell+1)^2}\right)-48 \left(\frac{1}{\ell+1}+\frac{1}{\ell}\right)+8 \left(\frac{1}{\ell+2}+\frac{1}{\ell-1}\right)\\\nonumber&
    +\frac{265933 }{1024}\left(\frac{1}{2 \ell+3}+\frac{1}{2 \ell-1}\right)+\frac{251475 }{1024}\left(\frac{1}{(2 \ell+3)^2}-\frac{1}{(2 \ell-1)^2}\right)\\\nonumber&+\frac{50625}{512} \left(\frac{1}{(2 \ell+3)^3}+\frac{1}{(2 \ell-1)^3}\right)
    +\frac{1225 }{8192}\left(\frac{1}{2 \ell+5}+\frac{1}{2 \ell-3}\right)-\frac{1473329}{4096 (2 \ell+1)}+\frac{120409}{128 (2 \ell+1)^3},\end{align}
    \begin{align}
    k_4^P &=576P\left(\frac{1}{\ell^5}-\frac{1}{(\ell+1)^5}\right)
-864P\left(\frac{1}{\ell^4}+\frac{1}{(\ell+1)^4}\right)+\left(4416P+32\right)
\left(\frac{1}{\ell^3}-\frac{1}{(\ell+1)^3}\right)\\\nonumber&
-\left(6472P+72\right)
\left(\frac{1}{\ell^2}+\frac{1}{(\ell+1)^2}\right)+\left(\frac{51044}{3}P-44\right)
\left(\frac{1}{\ell}-\frac{1}{\ell+1}\right)
-\frac{88}{3}P
\left(\frac{1}{(\ell-1)^2}+\frac{1}{(\ell+2)^2}\right)\\\nonumber
&+\left(\frac{1652}{9}P+\frac{196}{9}\right)
\left(\frac{1}{\ell-1}-\frac{1}{\ell+2}\right)-\left(1350P+\frac{3825}{32}\right)
\left(\frac{1}{(2\ell+3)^3}-\frac{1}{(2\ell-1)^3}\right)\\\nonumber
&-\left(3353P+\frac{57001}{192}\right)
\left(\frac{1}{(2\ell-1)^2}+\frac{1}{(2\ell+3)^2}\right)+\left(\frac{8653}{3}P-\frac{154543}{576}\right)
\left(\frac{1}{2\ell+3}-\frac{1}{2\ell-1}\right)\\\nonumber
&+\left(\frac{11}{12}P-\frac{3229}{2304}\right)
\left(\frac{1}{2\ell+5}-\frac{1}{2\ell-3}\right)+\frac{120409}{32(2\ell+1)^3}
-\frac{1473329}{1024(2\ell+1)}
+256\left(\frac{1}{\ell^3}+\frac{1}{(\ell+1)^3}\right)\\\nonumber
&+\frac{120409}{32(2\ell+1)^3}
-\frac{1473329}{1024(2\ell+1)}
+256\left(\frac{1}{\ell^3}+\frac{1}{(\ell+1)^3}\right)-576\left(\frac{1}{\ell^2}-\frac{1}{(\ell+1)^2}\right)
-192\left(\frac{1}{\ell}+\frac{1}{\ell+1}\right)
\\\nonumber
&+32\left(\frac{1}{\ell-1}+\frac{1}{\ell+2}\right)+\frac{50625}{128}
\left(\frac{1}{(2\ell-1)^3}
+\frac{1}{(2\ell+3)^3}\right),\end{align}
\begin{align}
k_5^P&=-1024 \left(\frac{1}{\ell^5}+\frac{1}{(\ell+1)^5}\right)
+3456 \left(\frac{1}{\ell^4}-\frac{1}{(\ell+1)^4}\right)
-8352 \left(\frac{1}{\ell^3}+\frac{1}{(\ell+1)^3}\right)\\\nonumber
&+16432 \left(\frac{1}{\ell^2}-\frac{1}{(\ell+1)^2}\right)
-\frac{230848}{3} \left(\frac{1}{\ell+1}+\frac{1}{\ell}\right)
-\frac{2768}{9} \left(\frac{1}{\ell+2}+\frac{1}{\ell-1}\right)\\\nonumber
&+\frac{176}{3} \left(\frac{1}{(\ell-1)^2}-\frac{1}{(\ell+2)^2}\right)
+\frac{237718850645}{6291456}
 \left(\frac{1}{2\ell+3}+\frac{1}{2\ell-1}\right)\\\nonumber
&+\frac{11443094731}{393216}
 \left(\frac{1}{(2\ell+3)^2}-\frac{1}{(2\ell-1)^2}\right)
+\frac{1349684225}{65536}
 \left(\frac{1}{(2\ell+3)^3}+\frac{1}{(2\ell-1)^3}\right)\\\nonumber
&+\frac{169745625}{16384}
 \left(\frac{1}{(2\ell+3)^4}-\frac{1}{(2\ell-1)^4}\right)
+\frac{11390625}{4096}
 \left(\frac{1}{(2\ell+3)^5}+\frac{1}{(2\ell-1)^5}\right)\\\nonumber
&+\frac{6281915}{1572864}
 \left(\frac{1}{2\ell+5}+\frac{1}{2\ell-3}\right)
+\frac{441}{2097152}
 \left(\frac{1}{2\ell+7}+\frac{1}{2\ell-5}\right)\\\nonumber
&-\frac{370685}{393216}
 \left(\frac{1}{(2\ell+5)^2}-\frac{1}{(2\ell-3)^2}\right)
+\frac{550779490991}{2359296(2\ell+1)}
+\frac{1244799693}{16384(2\ell+1)^3}
+\frac{41781923}{512(2\ell+1)^5},
\end{align}

\begin{align}
k_6^P=&-19200P\left(\frac{1}{\ell^7}-\frac{1}{(\ell+1)^7}\right)
+51840P\left(\frac{1}{\ell^6}+\frac{1}{(\ell+1)^6}\right)\\\nonumber
&-\left(234240P+512\right)
\left(\frac{1}{\ell^5}-\frac{1}{(\ell+1)^5}\right)
+\left(520800P+1728\right)
\left(\frac{1}{\ell^4}+\frac{1}{(\ell+1)^4}\right)\\\nonumber
&-\left(1435088P+4080\right)
\left(\frac{1}{\ell^3}-\frac{1}{(\ell+1)^3}\right)
+\left(\frac{124483808}{45}P+7856\right)
\left(\frac{1}{\ell^2}+\frac{1}{(\ell+1)^2}\right)\\\nonumber
&-\left(\frac{4221927584}{675}P+\frac{123812}{3}\right)
\left(\frac{1}{\ell}-\frac{1}{\ell+1}\right)
-\frac{2224}{9}P
\left(\frac{1}{(\ell-1)^3}-\frac{1}{(\ell+2)^3}\right)\\\nonumber
&+\left(\frac{90896}{45}P+\frac{4312}{27}\right)
\left(\frac{1}{(\ell-1)^2}+\frac{1}{(\ell+2)^2}\right)
-\left(\frac{246128}{27}P+\frac{73388}{81}\right)
\left(\frac{1}{\ell-1}-\frac{1}{\ell+2}\right)\\\nonumber
&+\frac{64}{225}
\left(\frac{1}{\ell-2}-\frac{1}{\ell+3}\right)
-\left(\frac{151875}{4}P+\frac{860625}{256}\right)
\left(\frac{1}{(2\ell+3)^5}-\frac{1}{(2\ell-1)^5}\right)\\\nonumber
&-\left(\frac{2263275}{16}P+\frac{12825225}{1024}\right)
\left(\frac{1}{(2\ell-1)^4}+\frac{1}{(2\ell+3)^4}\right)
-\left(\frac{40821881}{192}P+\frac{899841577}{36864}\right)
\left(\frac{1}{(2\ell+3)^3}-\frac{1}{(2\ell-1)^3}\right)\\\nonumber
&-\left(\frac{331670029}{1920}P+\frac{12381139373}{368640}\right)
\left(\frac{1}{(2\ell-1)^2}+\frac{1}{(2\ell+3)^2}\right)\\\nonumber
&+\left(\frac{4885477}{552960}-\frac{16643}{2880}P\right)
\left(\frac{1}{(2\ell-3)^2}+\frac{1}{(2\ell+5)^2}\right)
+\left(\frac{1031179}{172800}P-\frac{597735581}{33177600}\right)
\left(\frac{1}{2\ell+5}-\frac{1}{2\ell-3}\right)\\\nonumber
&+\left(\frac{31}{460800}P-\frac{108347}{29491200}\right)
\left(\frac{1}{2\ell+7}-\frac{1}{2\ell-5}\right)-\left(\frac{45994593719}{55296}P+\frac{474662050279}{10616832}\right)
\left(\frac{1}{2\ell+3}-\frac{1}{2\ell-1}\right),
\end{align}
where we defined
\begin{align}
    \psi^{(n)}_+(\ell) &= \psi ^{(n)}(\ell-1)-2 \psi ^{(n)}(\ell)+3 \psi ^{(n)}(\ell+1)-2 \psi ^{(n)}(\ell+2)+\psi ^{(n)}(\ell+3),\\
    \psi^{(n)}_-(\ell)&= 2 \psi ^{(n)}(\ell) -3 \psi ^{(n)}(l+1)+2 \psi ^{(n)}(\ell+2).
\end{align}

\subsection{Exponentiated amplitude}
\noindent We write the following result in terms of $x$ using $z(x)=1-2x^2$. Additionally, the master integrals are represented with $I^j_{\vec{g}}(x)$ being the $j$th component of $\vec{I}_{\vec{g}}(x)$ ($j=1,...,6$), and using the compressed HPL notation for $\vec{g}$ such that, for example, $I^j_{\{3,-2,1\}}=I^j_{\{0,0,1,\,0,-1,\,1\}}(x)$.
The EFT exponentiated amplitude through $\mathcal{O}(G^7)$ is then
\begin{equation}\label{eq:DeltaHresult}
\setlength{\jot}{8pt}
\begin{split}
\Delta^H(z(x),\omega)&=\frac{GM\pi}{8x^2}\\
&\quad +(GM)^3\omega^2\pi\left[\frac{p_1(x)}{3x^8}\left(H_{\{1,0\}}(x)-H_{\{-1,0\}}(x)\right)+\frac{p_2(x)H_{\{0\}}(x)}{9x^6}-\frac{p_3(x)}{108x^6}\right]\\
&\quad+(GM)^5\omega^4\pi\left[-\frac{\lambda^{(-)}_{0,2}-\lambda^{(+)}_{0,2}}{64\pi (GM)^5 M_\text{Pl}^2}+\frac{p_4(x)}{1620x^6}-\frac{p_5(x)H_{\{0\}}(x)}{135x^6}+\frac{p_6(x)}{45x^8}\left(H_{\{-1,0\}}(x)-H_{\{1,0\}}(x)\right)\right.\\
&\qquad\qquad\left.+\frac{32p_7(x)}{15x^6}\left(I^4_{\{3\}}(x)+I^5_{\{3\}}(x)\right)+\frac{32p_8(x)}{5x^8}\left(I^5_{\{4\}}(x)-I^4_{\{4\}}(x)\right)\right]\\
&\quad+(GM)^6\omega^5\pi\left[-\frac{\lambda^{(-)}_{0,2}-\lambda^{(+)}_{0,2}}{32(GM)^5M_\text{Pl}^2}-\frac{i\left(\lambda^{(-)}_{1,2}-\lambda^{(+)}_{1,2}\right)}{32\pi(GM)^5 M_\text{Pl}^2}\right]\\
&\quad+(GM)^7\omega^6\pi\left[\frac{1}{\epsilon}\left(\frac{\lambda^{(-)}_{2,2}\left(\epsilon^{-1}\right)-\lambda^{(+)}_{2,2}\left(\epsilon^{-1}\right)}{16 \pi  (GM)^5 M_\text{Pl}^2}-\frac{107 \left(\lambda^{(-)}_{0,2}-\lambda^{(+)}_{0,2}\right)}{6720 \pi (GM)^5 M_\text{Pl}^2}\right)\right.\\
&\qquad\qquad\left.+\frac{(6x^2-5)\left(\lambda^{(-)}_{3,0}-\lambda^{(+)}_{3,0}\right)}{576\pi(GM)^7 M_\text{Pl}^2}+\frac{\lambda^{(-)}_{2,2}-\lambda^{(+)}_{2,2}}{16\pi(GM)^5 M_\text{Pl}^2}-\frac{i\left(\lambda^{(-)}_{1,2}-\lambda^{(+)}_{1,2}\right)}{16(GM)^5 M_\text{Pl}^2}\right.\\
&\qquad\qquad\left.-\frac{\left(\lambda^{(-)}_{0,2}-\lambda^{(+)}_{0,2}\right)}{3360\pi(GM)^5 M_\text{Pl}^2}\left(107\left(2\log\left(\frac{\mu}{\omega}\right)+\log(\mu)\right)+420\zeta_2\right)+\frac{\lambda^{(-)}_{2,0}p_9(x)}{1411200\pi(GM)^5 M_\text{Pl}^2x^6}\right.\\
&\qquad\qquad\left.+\frac{\lambda^{(+)}_{2,0}p_{10}(x)}{2822400 \pi (GM)^5M_\text{Pl}^2x^2(x^2-1)}-\frac{\lambda^{(-)}_{2,2}(\epsilon^{-1})p_{11}(x)}{96\pi(GM)^5 M_\text{Pl}^2x^6}-\frac{\lambda^{(+)}_{2,2}(\epsilon^{-1})p_{12}(x)}{96\pi(GM)^5 M_\text{Pl}^2x^2(x^2-1)}\right.\\
&\qquad\qquad\left.-\frac{p_{13}(x)}{85050x^6(x^2-1)}+\frac{p_{14}(x)H_{\{0\}}(x)}{567x^6}+\frac{p_{15}(x)}{189x^8}\left(H_{\{1,0\}}(x)-H_{\{-1,0\}}(x)\right)\right.\\
&\qquad\qquad\left.-\frac{p_{16}(x)}{63x^6}\left(I^4_{\{3\}}(x)+I^5_{\{3\}}(x)\right)-\frac{p_{17}(x)}{7x^6}\left(I^4_{\{5\}}(x)+I^5_{\{5\}}(x)\right)+\frac{p_{18}(x)}{63x^8}\left(I^4_{\{4\}}(x)-I^5_{\{4\}}(x)\right)\right.\\ 
&\qquad\qquad+\frac{p_{19}(x)}{7x^8}\left(I^4_{\{6\}}(x)-I^5_{\{6\}}(x)\right)\Bigg]
\end{split}
\end{equation}
with
\begin{equation}
\begin{split}
p_1(x)&=(x^2-2)(x^4-10x^2+10)\\
p_2(x)&=11x^4-60x^2+60\\
p_3(x)&=121x^4-450x^2+360\\
p_4(x)&=96979x^4-1279170x^2+1620360\\
p_5(x)&=8093x^4-166200x^2+270060\\
p_6(x)&=1017x^6-33792x^4+114090x^2-90020\\
p_7(x)&=11x^4-528x^2+1140\\
p_8(x)&=x^6-120x^4+732x^2-760\\
p_9(x)&=12x^6\left(3745\gamma_E-30417-3745\log(\pi)\right)-18725x^2+11235\end{split}\end{equation}
\begin{equation}\begin{split}
p_{10}(x)&=7490+x^2(x^2-1)\left(-89880\gamma_E+619933+89880\log(\pi)\right)\\
p_{11}(x)&=12x^6\log(\mu)+8x^6+5x^2-3\\
p_{12}(x)&=-2x^4+12(x^2-1)x^2\log(\mu)+2x^2+1\\
p_{13}(x)&=4842x^8+403242233x^6-15523299785x^4+37366876470x^2-22246823970\\
p_{14}(x)&=2\left(1268857x^4-77799036x^2+148312164\right)\\
p_{15}(x)&=2\left(167679x^6-16005900x^4+60036186x^2-49437388\right)\\
p_{16}(x)&=128\left(1133x^4-160209x^2+391020\right)\\
p_{17}(x)&=256\left(11x^4-4356x^2+13164\right)\\
p_{18}(x)&=128\left(569x^6-136653x^4+770676x^2-782040\right)\\
p_{19}(x)&=512\left(x^6-1092x^4+10938x^2-13164\right)
\end{split}
\end{equation}
$\Delta^V(z(x),\omega)$ and the full amplitude $\A^I(z(x),\omega)$ are provided in the ancillary Mathematica file attached to this work.

\subsection{Summary of the ancillary file}
\noindent We provide lengthy results in the attached Mathematica file \texttt{EFTAmplitudeData.wl}. The file contains
\begin{itemize}
    \item The EFT graviton partial-wave phase shifts $\delta_\ell^P(\omega)$ at $\ell=2$ and $\ell=3$, with parity $P=\pm$, through $\mathcal{O}(G^7)$: \texttt{deltaPLUSl2G7} (Eq. \eqref{eq:deltaPLUSl2G7}), \texttt{deltaMINUSl2G7}  (Eq. \eqref{eq:deltaMINUSl2G7}), \texttt{deltaPLUSl3G7}, \texttt{deltaMINUSl3G7}.
    \item The EFT exponentiated amplitude $\Delta^I(z(x),\omega)$, for $I=H,V$, given separately at each order in $G$ through $\mathcal{O}(G^7)$: \texttt{EFTDeltaH1}, \texttt{EFTDeltaV1}, ..., \texttt{EFTDeltaH7}, \texttt{EFTDeltaV7}. Therefore Eq.~\eqref{eq:DeltaHresult} is 
    \begin{equation}\nonumber
        \Delta^H(z(x),\omega)=\sum_{n=1}^7\texttt{EFTDeltaHn}
    \end{equation}
    \item The EFT full amplitude $\A^I(z(x),\omega)$ (Eq.~\eqref{eq:PW}), for $I=H,V$, given separately at each order in $G$ through $\mathcal{O}(G^7)$: \texttt{EFTAmpH1}, \texttt{EFTAmpV1}, ..., \texttt{EFTAmpH7}, \texttt{EFTAmpV7}.
\end{itemize}
The results are expressed in terms of the master integrals $I^j_{\vec{g}}(x)$ defined in Eq.~\eqref{eq:masterInts}, with $j=1,\ldots,6$. In the ancillary file, an integral is represented as \texttt{MI[j,\{g1,...,gk\}]}, where the indices use the compressed HPL convention. For example, $\texttt{MI[j,\{3,-2,1\}]}=I^j_{\{0,0,1,\,0,-1,\,1\}}(x)$.

The Love numbers use the following notation: 
\begin{align}
    \lambda^{(+)}_{n,\ell}&=\texttt{\textbackslash[Lambda]E[n,l]}\nonumber\\
    \lambda^{(-)}_{n,\ell}&=\texttt{\textbackslash[Lambda]B[n,l]}\nonumber\\
    \lambda^{(+)}_{n,\ell}(\epsilon^{-1})&=\texttt{\textbackslash[Lambda]E[n,l,-1]}\nonumber\\
    \lambda^{(-)}_{n,\ell}(\epsilon^{-1})&=\texttt{\textbackslash[Lambda]B[n,l,-1]}\nonumber
\end{align}

\vspace{-0.9cm}
\twocolumngrid

\end{document}